# Clustering in VANET: Algorithms and Challenges

Mohammad Mukhtaruzzaman and Mohammed Atiquzzaman

*School of Computer Science, University of Oklahoma,* Norman, OK-73019

mukhtar@ou.edu, atiq@ou.edu

*Abstract*—**Clustering is an important concept in vehicular ad hoc network (VANET) where several vehicles join to form a group based on common features. Mobility-based clustering strategies are the most common in VANET clustering; however, machine learning and fuzzy logic algorithms are also the basis of many VANET clustering algorithms. Some VANET clustering algorithms integrate machine learning and fuzzy logic algorithms to make the cluster more stable and efficient. Network mobility (NEMO) and multi-hop-based strategies are also used for VANET clustering. Mobility and some other clustering strategies are presented in the existing literature reviews; however, extensive study of intelligence-based, mobility-based, and multi-hop-based strategies still missing in the VANET clustering reviews. In this paper, we presented a classification of intelligence-based clustering algorithms, mobility-based algorithms, and multi-hop-based algorithms with an analysis on the mobility metrics, evaluation criteria, challenges, and future directions of machine learning, fuzzy logic, mobility, NEMO, and multi-hop clustering algorithms.**

## 1 INTRODUCTION

Vehicular communication for intelligent transport systems (ITS) is a rapidly growing research area. Wireless access in vehicular environments (WAVE) is defined for wireless communication in vehicular ad hoc network (VANET) on the dedicated short-range communications (DSRC) frequency bands by IEEE802.11P and IEEE1609 [1]. DSRC/WAVE is currently used to satisfy the low latency requirement for safety and control messages for vehicle-to-vehicle (V2V) communication and long-term evolution (LTE) is used for vehicle to infrastructure (V2I) communication. 5G millimeter waveband is under research to provide ultra-low-latency for vehicular communication. During V2V communication, each vehicle acts as a mobile router and an on-board unit (OBU) is used in each vehicle to communicate with the other vehicles. In this paper, vehicle, car, and node are used interchangeably to mean vehicle.

VANET has some common features with mobile ad hoc network (MANET); however, VANET has its unique features, such as high mobility that differentiates it from MANET. The vehicles in VANET do not suffer from energy deficiency but faces many new challenges due to their high mobility. When the number of vehicles increases, scalability becomes an important issue. In the absence of any central infrastructure, VANET suffers high packet loss due to a large volume of message dissemination among the vehicles for V2V communication. VANET also suffers from issues such as the hidden terminal problem, high latency for safety message transmission, message security, broadcast storm problem, quality of service (QoS), packet routing, congestion control, and resource management. To solve these issues, a hierarchical structure has been investigated in the literature [2]. In a hierarchical structure, two or more nearby vehicles, who have some common features, join in a group which is called clustering. Clustering concept is widely used in data mining, and machine learning [3]. Clustering is also used in MANET, which is the predecessor of VANET, to cluster the mobile nodes. In a clustered vehicular environment, a large network of vehicles is considered as a network of some small networks or clusters.

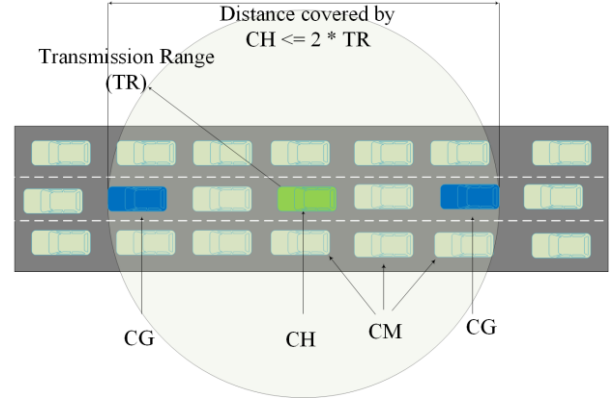

Fig. 2. Different components of VANET clustering.

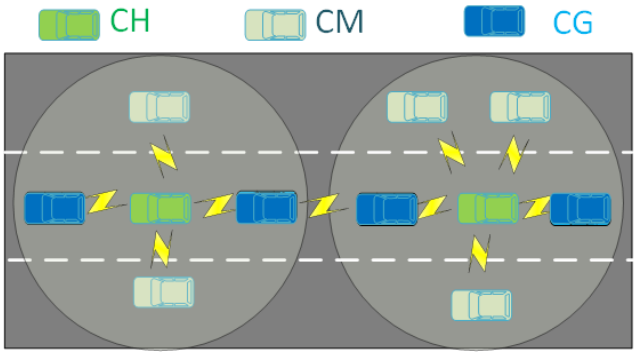

Fig. 1. Basic clustering concept in VANET.

### 1.1 Clustering Concept in VANET

In VANET clustering, cluster head (CH) plays a key role in the formation process of a cluster, as shown in Fig. 1. A cluster can be created in various ways based on the input metrics. The member vehicle of a cluster is called cluster member (CM). Other than CH and CM, some algorithms use two CMs to communicate with other clusters on behalf of the CH are called cluster gateways (CGs). Unless specified as CG, all members of a cluster are termed as CMs. One CH, zero/one/two CGs, and any number of CMs can be present in a cluster. In VANET clustering, CH acts like a mobile router and CM acts like a mobile node. The role of CG lies between CH and CM. The cluster is formed based on the metrics such as the average relative velocity of the vehicle, acceleration, position, direction, the degree of the vehicle, the density of the vehicles, transmission range, etc. CH is selected from the vehicles which is most stable among the participating vehicles. The rest of the vehicles join the cluster as CMs. Therefore, CH selection is a part of the cluster formation process and no separate CM selection criteria need to be presented. CH and CMs maintain a routing table containing information of the CH and CMs of the cluster for intra-cluster communication. However, CM does not maintain any routing table for other clusters, which is maintained by the CH, if necessary. Hence, a large network is considered as a group of some small networks or clusters.

The coverage of a cluster is limited by the transmission range (TR) of the CH, as shown in Fig. 2. Since the distance covered by a CH is limited by its TR, a vehicle which is located at the edge of a cluster has a high probability of losing connection with the CH. The relative speed of two vehicles can vary all the time depending on the speed of two vehicles. When a vehicle's position is at the edge of a cluster, the vehicle may enter and exit the TR of the CH frequently due to the change in relative speed and the vehicle will lose connection with the CH more frequently. As a result, data loss will be very high when a vehicle remains at the edge of a cluster. For this reason, some algorithms

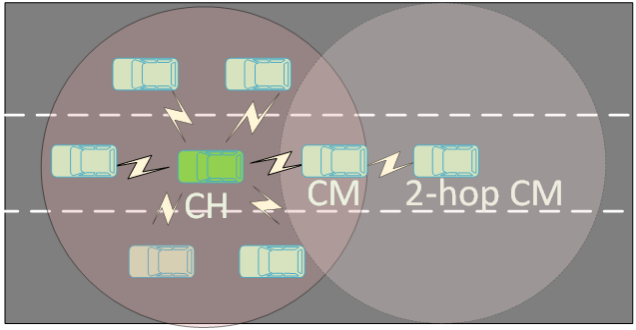

Fig. 3. Multi-hop clustering in VANET.

prefer a geographically center vehicle as the CH for reliability.

### 1.2 Multi-hop Clustering

Generally, a cluster of vehicles means a 1-hop cluster where a CH can reach all its CMs directly because the CMs are within the range of the CH; however, some clustering algorithms are based on multi-hop strategy. When a vehicle cannot reach the CH of a cluster directly but can reach a member of the cluster, then the new vehicle joins to the cluster through a CM. Hence, a CH can cover CMs in a multi-hop manner which is termed as multi-hop clustering, or $N$-hop clustering, or $k$-hop clustering. The value of $N$ or $k$ depends on the number of hops the CH can cover. In the Fig. 3, the 2-hop CM cannot reach the CH but can reach a CM of the CH. As a result, the 2-hop CM joins the cluster through a CM of the CH creating a multi-hop cluster.

### 1.3 Objective and Motivation

In machine learning and data mining, many algorithms have been developed for efficient clustering. Many VANET clustering techniques are based on machine learning algorithms such as k-means clustering, hierarchical clustering, etc. Another domain for VANET clustering is fuzzy logic where future movement of the vehicle is predicted using fuzzification and defuzzification.

Many hybrid architectures are also proposed for VANET clustering where machine learning-based algorithms are integrated with fuzzy logic to create efficient and stable clusters by selecting a more suitable vehicle as the CH. Previous surveys (see Section 1.4) on VANET clustering did not take into account intelligence-based clustering algorithms such as machine learning and fuzzy logic-based algorithms. Besides intelligence-based strategies, many algorithms emphasize the mobility parameters to provide stable clusters. Network mobility (NEMO) concept is also used for vehicular communication since NEMO has similarity with the clustering technique where the mobile router (MR) moves from one place to another place with its mobile nodes (MNs) and MNs communicate through MR only. This scenario can be compared with the CH and CMs in the clustering algorithm. Hence, MR, MN, and access router (AR) in NEMO are equivalent to CH, CM, and RSU in the VANET. Additionally, multi-hop clustering strategies are also used in VANET clustering where the CH can cover more than one hop area to reduce the number of clusters. Therefore, our *aim* is to classify all three major types of VANET clustering algorithms: intelligence, mobility, and multi-hop to stimulate the research of efficient and stable clustering algorithms for VANET.

### 1.4 Literature Survey on Clustering

The first attempt to study clustering algorithms for VANET is [4]; however, only a few clustering techniques have been presented without any classification. The first classification of the VANET clustering techniques is presented in [5]. Classification is performed based on position, destination, and medium access, etc. Few of the algorithms presented are based on intelligence, vehicle mobility, and multi-hop strategies; however, the absence of NEMO strategies and presentation of a very few intelligence and multi-hop algorithms without any further classification are not enough to study VANET clustering. Some mobility-based clustering techniques are presented in [6] along with ID-based, degree-based, direction-based algorithms, etc. Some machine learning and fuzzy logic-based strategies are discussed in [7] without classification.

TABLE I
COMPARISON OF EXISTING SURVEYS ON VANET CLUSTERING

| Ref. | Machine learning | Fuzzy logic | Vehicle mobility | Network mobility | Multi-hop | Comments |
|------|------------------|-------------|------------------|------------------|-----------|----------|
| [4]  | -  | -  | -  | -  | -  | Some clustering techniques without classification |
| [5]  | √  | -  | √  | -  | √  | Limited number of ML and multi-hop algorithms |
| [6]  | -  | -  | √  | -  | -  | Focused on Mobility issue |
| [7]  | √  | -  | -  | -  | -  | Some Machine learning clustering are discussed |
| [8]  | -  | -  | -  | -  | -  | Discussed one paper from different types of clustering |
| [9]  | -  | -  | √  | -  | -  | Detail classification is presented based on application and CH selection criteria |
| [10] | √  | -  | -  | -  | -  | Some machine learning clustering techniques are discussed |
| Our Survey | √ | √ | √ | √ | √ | Machine learning-based, fuzzy logic-based, and NEMO-based algorithms are presented along with mobility and multi-hop-based algorithms |

In [8], beacon message, density, direction, etc. are considered to classify the existing clustering approaches; however, only a single technique is discussed from each group. A detail classification of VANET clustering is presented in [9] based on the cluster application where a flow of the clustering techniques starting from its MANET origins is discussed. CH selection criteria, CG and CM selection metrics, etc. are described in detail with a discussion on simulators used for VANET clustering. A detail description of mobility-based clustering strategies is presented; however, no classification is presented for intelligence-based or multi-hop-based strategies. Machine learning concepts are described in [10] for the vehicular environment including few machine learning clustering techniques. A comparison of the existing surveys is shown in Table I.

Some of the review papers discussed the machine learning-based strategies in a narrow scope while none of them give any concentration on fuzzy logic or hybrid strategies of machine learning and fuzzy logic. Vehicle mobility is covered by some papers while neglected the NEMO issue. Some reviews included multi-hop strategies but lack any detail classification.

In VANET clustering, many fuzzy logic-based algorithms have been proposed along with machine learning-based algorithms. Some NEMO algorithms are also used for VANET clustering along with vehicle mobility-based algorithms. Moreover, many multi-hop strategies have been proposed for VANET clustering to reduce the number of clusters. Therefore, we need to study all these research works extensively to study VANET clustering comprehensively, which is absent in the literature survey. During the classification of the VANET clustering algorithms, all the existing papers lack some important strategies such as machine learning, fuzzy logic,

NEMO, and multi-hop algorithms. Therefore, a comprehensive study of machine learning, fuzzy logic, mobility, NEMO, and multi-hop strategies is still not present in the literature. Hence, we get three broader categories to classify VANET clustering algorithms: intelligence, mobility, and multi-hop-based clustering algorithms.

*1.5 Contributions*

The *contributions* of this paper can be summarized as follows:

1. The paper studied intelligence-based VANET clustering extensively, classifying them into machine learning-based and fuzzy logic-based along with a comparison among the algorithms in terms of strengths and weaknesses.
2. Comparison among hybrid architectures which combine machine learning and fuzzy logic algorithms to exploit the advantages of both the schemes presented.
3. Apart from vehicle mobility-based strategies, network mobility-based strategies studied separately.
4. Details study of multi-hop strategies presented in a separate section.

*1.6 Structure*

The rest of the paper is organized as follows. In Section 2, we presented the classification of VANET clustering. Intelligence-based VANET clustering algorithms are presented in Section 3. Mobility-base algorithms are classified in Section 4, where multi-hop-based algorithms are presented in Section 5. Present challenges and future research directions are presented in Section 6 with a conclusion in Section 7.

**2 TAXONOMY OF VANET CLUSTERING**

Based on the algorithms, VANET clustering schemes can be single-hop or multi-hop. The single-hop strategies can be divided into two larger groups based on their algorithms: intelligence-based strategies and mobility-based strategies. Hence, VANET clustering schemes can be divided into three main categories as described in Section 1.3 and Section 1.4. Therefore, we classified the clustering schemes in VANET into three categories: intelligence-based strategies, mobility-based strategies, and multi-hop-based strategies as shown in Fig. 5. Intelligence based strategies are further classified into machine learning, fuzzy logic, and hybrid algorithms. Mobility-based strategies are divided into vehicle mobility and network mobility algorithms while multi-hop-based strategies are divided into 2-hop and 2+ hop algorithms based on the number of hop count. Each group of algorithms is further classified in the sub-group, which is not shown in Fig. 5, but presented in Sections 3, 4, and 5.

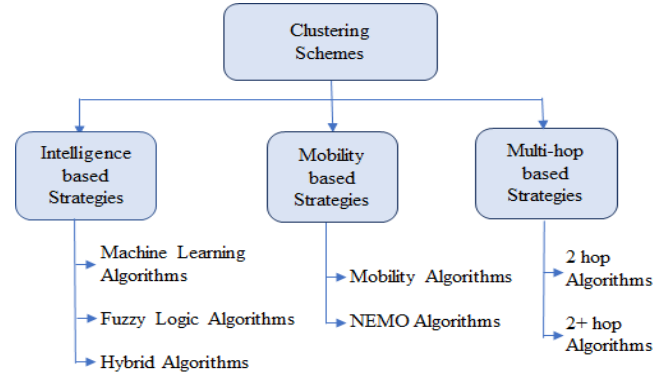

Fig. 4. Taxonomy of VANET clustering schemes.

Evaluation criteria of the algorithms presented in Tables II through IX in Sections 3 through 5. A summary of the algorithms presented at the end of each section where overall challenges and future directions of VANET clustering algorithms presented in Section 6.

## 3 INTELLIGENCE BASED STRATEGIES

Clustering is an important concept in machine learning and data mining [3] and many clustering algorithms are developed over the years such as k-means and hierarchical clustering. The clustering algorithms from machine learning are used in VANET for vehicle clustering. Fuzzy logic is also used for VANET clustering. Supervised learning, such as Q-learning, and other machine learning clustering algorithms are used along with fuzzy logic to create a hybrid strategy for VANET. The *difference* between our works with the other works is, we classified and analyzed machine learning algorithms and fuzzy logic-based algorithms separately. We also analyzed the hybrid architecture of machine learning and fuzzy logic in a separate section. In

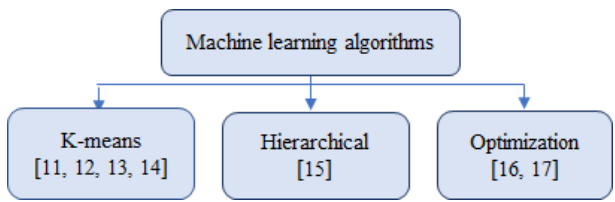

Fig. 5. Classification of Machine learning-based VANET clustering algorithms.

Section 3.1, we discussed machine learning algorithms for VANET clustering strategies. Next, fuzzy logic algorithms described in Section 3.2 and hybrid clustering strategies placed in Section 3.3. An analysis of intelligence-based strategies presented in Section 3.4.

### 3.1 Machine Learning Algorithms

Clustering algorithms are used in data mining and machine learning to cluster similar types of objects. K-means algorithm is the most frequently used machine learning algorithm in VANET where *k* number of clusters are created dividing the vehicles. Initial centroids are assumed and coordinates of the vehicle are given as input. In the next step, Euclidean distances are calculated to determine the new centroid, and the centroids are elected as the CH. Whenever a CM joins into a cluster, or leave a cluster, the mean of the cluster is susceptible to change in k-means algorithm and need to re-calculate the new mean of the cluster to reflect the change that can lead to elect a new vehicle as the CH. In hierarchical clustering, Euclidean distances of all vehicles are calculated and the vehicles connect with each other sequentially starting from the minimum distance.

#### 3.1.1 K-means Clustering Algorithms

A k-means-based clustering algorithm is proposed by Bansal et al. [11] to divide the vehicles into clusters. Three parameters: x dimension and y dimension, i.e., the position of the vehicles is considered to form the clusters. The number of clusters is given as input, then a modified k-means algorithm is applied to divide the vehicles into clusters. To choose the CH, the centroid of the cluster is selected along with some security issues. To increase security, a hashing technique is used to encrypt or decrypt the packets. After selecting the centroid as the CH, the rest of the vehicles join the cluster as the CMs of the cluster. No separate maintenance phase is required since the clusters cannot overlap in intelligent clustering including k-means algorithm. Simulation results show that PDR and throughput can be improved in the proposed algorithm while routing overhead increases compare to base k-means algorithm; however, the number of clusters is an input of this algorithm, but density and number of vehicles can vary in a different scenario. Hence, the number of clusters should be an independent variable that can increase or decrease depending on the number of vehicles and density. Otherwise, the number of CMs in a cluster can be very high or very low. Additionally, if any vehicle joins or leaves the cluster, the mean of the entire cluster can be changed with the change of the CH itself that reduces the cluster lifetime and cluster stability.

Instead of enhancing security measure as in [11], k-means is used to solve data congestion problem [12] to decrease packet loss and end-to-end delay. Clustering is performed using k-means algorithm based on distance and direction along with message size, the validity of messages, and type of messages. Two types of control strategies have been used: open-loop and closed-loop solutions. The open-loop solutions prevent congestion before it happens while closed-loop solutions control the congestion after detection. Instead of vehicles, clustering of messages is performed at RSU where features, number of clusters and number of iterations are given as input. However, the number of clusters is fixed, and initial centroids are set based on a first come first serve basis which is inefficient for stability and cluster lifetime.

K-means [13] is used to increase the stability of the clusters. The distances of the vehicles are calculated to find the minimum average distance to form a cluster. The center vehicle is selected as the CH. Distance is measured by Euclidean distance and all pair shortest path is calculated within a cluster using the Floyd-Warshall algorithm to choose the CH. However, the limitation of the number of clusters persists, as was in [11-12].

One limitation is common for [11-13], which is sensitivity to the initial centroids. To overcome this drawback adaptive k-harmonic means is proposed [14] where a vehicle must meet the minimum bandwidth requirement to be elected as the CH. Traditional k-harmonic means, where relative distance and centroids are measured, is modified to make compatible with the mobility of vehicles. The velocity of the vehicles is considered along with their position to form the clusters. However, the limitation of the fixed number of clusters is continued that can cause a problem in v2v communication with many vehicles.

### 3.1.2 Hierarchical Clustering Algorithms

To overcome the limitations of the k-means algorithm, an agglomerative hierarchical clustering approach is used in [15], where the direction and speed of the vehicles are considered to form a cluster along with some quality of service (QoS) parameters. The past duration of the node acting as a CH, PDR, and TR are considered for CH. In hierarchical clustering, Euclidean distances of all the vehicles are calculated and the vehicles connect with each other sequentially starting from the minimum distance and do not require the number of clusters as the input unlike k-means [11-14]. The vehicles in [15] are considered as two clusters based on their direction and CH is selected based on the duration of acting as a CH in the past. However, while the implementation of k-means is simple, the hierarchical approach requires a proximity matrix calculation which leads to $O(n^2)$ space complexity and $O(N^2 \log(n))$ to $O(N^3)$ time complexity. Moreover, once a CM joins to a cluster, it cannot be undone, but topology can change any time in VANET and requires a change in the cluster also.

### 3.1.3 Optimization Algorithms

A different approach than [11-15], a nature-inspired algorithm is used in [16] and [17] based on moth-flame optimization (MFO). The MFO algorithm depends on the navigation method of moths, which follow a spiral flying path, called transverse orientation. Moth can fly maintaining a fixed angle with the moon that can be considered as a straight path for a long distance. The same concept is used [16, 17] to cluster the vehicles in a highway environment to optimize cluster considering speed, direction, grid size, the degree of a node, and transmission range of the vehicles, however, no result is provided to prove the optimization of performances in terms of PDR, end-to-end delay, number of hops, and throughput.

TABLE II
EVALUATION OF MACHINE LEARNING-BASED STRATEGIES

| Ref. | Algorithm | Evaluation parameters |
|---|---|---|
| [11] | K-means | PDR, throughput, overhead |
| [12] | K-means | PDR, throughput, delay |
| [13] | K-means | CH Duration, signal quality, TR |
| [14] | K-Harmonic means | Coverage |
| [15] | Hierarchical | PDR, throughput, delay, degree of node |
| [16] | MFO | No. of Clusters, TR |
| [17] | MFO | No. of Clusters, grid size |

Table II presents the VANET clustering algorithms which are based on machine learning clustering algorithms.

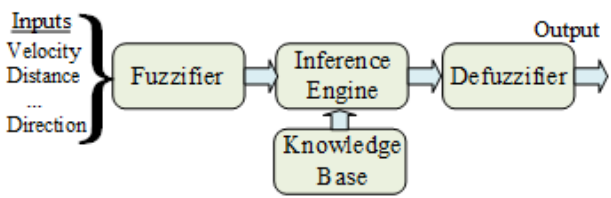

Fig. 6. Structure of Fuzzy logic system in VANET.

### 3.2 Fuzzy Logic Algorithms

Many clustering strategies in VANET are based on fuzzy logic algorithms. Instead of the value as true or false, the degree of certainty is considered in the fuzzy logic system (FLS). The steps of FLS are shown in Fig. 6. Five steps of FLS can be considered in terms of VANET. In the first step, the input parameters such as relative speed, vehicle distance, moving direction, and acceleration are defined. In the second step, fuzzification is performed where a fuzzifier transforms the input parameters into a fuzzy set. The third step is performed by an inference engine where the fuzzy rules are defined based on the knowledge base and applied on the fuzzy set to produce the output fuzzy sets. Defuzzification process is performed by a defuzzifier in the next phase to generate crisp output values from the output fuzzy sets. In the last step, tuning of the system is performed reviewing the range of the inputs and outputs, revising the fuzzy sets, and tuning the rules. Classification of fuzzy logic algorithms is presented in Figure 7.

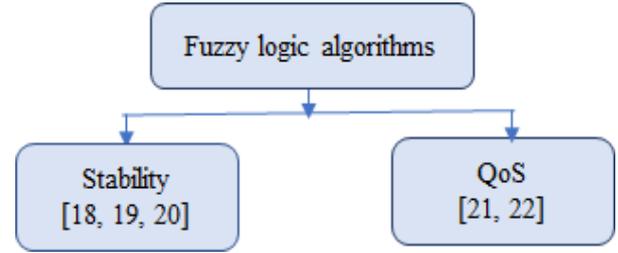

Fig. 7. Classification of FL-based clustering.

#### 3.2.1 Stability Algorithms

A fuzzy logic-based CH selection algorithm is proposed by Hafeez et al. [18], the first instance of introducing fuzzy logic system in VANET scenario. In this earlier work, relatively simple fuzzy system developed where two metrics such as relative speed and distance are given as input to the fuzzifier to start the clustering formation process. The fuzzy logic inference system is used to learn the driver's behaviors, and to predict future speed and position. Based on the predicted speed and position, the CH is selected by the defuzzifier. If the stability factor of the CH falls below a predefined threshold value, a new member is selected as the CH. Merging of two clusters are allowed in this scheme. If the second CH reaches to half of the TR of the first CH, the second CH will merge with the first CH; however, considering only two input parameters in the fuzzy input sets affects the performance of the selection of the CH.

The work in [18] is improved in [19] adding acceleration as the input parameter along with speed and distance to create more stable clusters. Fuzzy logic inference system is integrated with an adaptive learning mechanism to provide a more stable cluster by predicting the future speed and the positions of the CMs. Like [18], the stability of the vehicle compared to the neighbor vehicles is given preference to be selected as the CH. Similarly, merging of two clusters are allowed and follow the same process as in [18]; however, three input parameters are also proved insufficient for the highly dynamic nature in VANET. To further improve the performance of [19], the direction of the vehicles is also considered [20] along with the speed, distance, and acceleration. However, previous history of acting as CH and QoS issues are not considered in this approach.

### 3.2.2 QoS Algorithms

Stability is given priority in [18-20] without considering QoS issue. In [21], a hybrid network architecture of V2V and LTE advanced cellular network is proposed where quality-of-service is improved using a fuzzy logic-based gateway selection technique. Cluster is formed considering traffic type of the vehicle and the CH is selected using received signal strength and load. The CH is the leader of the cluster but may not work as the gateway to communicate with the LTE advanced cellular network [21] which is a unique concept for RSU assisted VANET clustering strategies while CH or CG is generally selected as the LTE gateway in VANET. However, dynamic clustering at higher speed will cause frequent CH change as well as frequent change of CG candidates and will increase the complexity selecting the gateway node that can increase packet loss and end-to-end delay.

To solve the limitation of resources in dynamic vehicular cloud architecture, a fuzzy-based CH selection process is proposed [22]. To improve reliability and QoS, a CH works as a cloud controller who can create, delete and update the vehicular cloud. The average speed, degree of node, and link quality are considered to form the clusters and the CH is selected based on a fit factor; however, reliability and quality-of-service of such strategy is questionable, because the increase in degree of node can decrease the performance of the CH by allocating resources to a large number of members. Moreover, performance of the scheme degraded in the absence of RSU.

Table III presents the VANET clustering algorithms which are based on fuzzy logic algorithms.

### 3.3 Hybrid Algorithms

Machine learning algorithms are integrated with fuzzy logic system to make the cluster formation process and CH selection process more efficient in a hybrid manner.

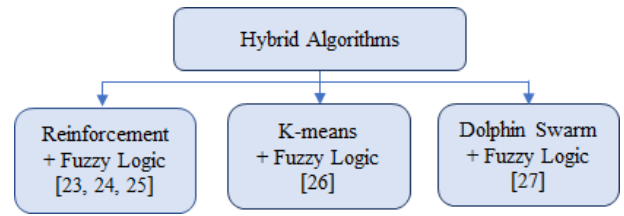

Fig. 8. Classification of hybrid of Machine learning and Fuzzy logic clustering algorithms.

### 3.3.1 Reinforcement Learning Algorithms with Fuzzy Logic

To improve efficiency, stability, and reliability of cloud services in a vehicular environment, a new architecture is proposed [23] using a reinforcement learning algorithm, Q-learning, along with the fuzzy logic algorithm used in [22]. The CH is selected based on fuzzy logic and resource management is improved using Q-learning based service provider selection technique. CMs are limited to the communication range of the CH. CH is selected based on the fit factor, like [22], where the cluster is formed depending on speed, degree, and RSU link quality. Every vehicle broadcasts its fit factor to be selected as the CH. Resource management is improved from [22] by deploying the Q-learning technique to select the service provider from the neighborhood vehicles that improve the efficiency of the CH selection process. Three different queuing methods such as first in first out, bandwidth aware, and resource-aware are used [23] that were not present in [22]. However, among the mobility parameters, the relative speed is only considered to select the CH, other parameters such as acceleration and direction are ignored. Moreover, the algorithm is RSU dependent.

TABLE III
EVALUATION OF FUZZY LGIC-BASED STRATEGIES

| Ref. | Fuzzy Input | Evaluation |
|---|---|---|
| [18] | Relative Speed, Distance | CH duration, cluster size, CM duration |
| [19] | Speed, distance, acceleration | CH duration, cluster size, CM duration, delay, reliability, cluster size, PDR |
| [20] | Speed, acceleration, direction | CH candidate values |
| [21] | Position, velocity | Delay, packet loss, throughput |
| [22] | Speed, degree of node, link quality | CH duration, service delay |

A data storage scheme is proposed [24] that store the data employing a fuzzy logic-based protocol considering multiple metrics such as throughput, stability, and bandwidth efficiency. To increase the stability of the fuzzy decision, Q-learning is used. The slow vehicles are selected as the CH to avoid frequent change of cluster heads to make the cluster more stable considering vehicle velocity, degree of node, and channel condition. However, slow vehicles cannot be the most suitable candidate to become a CH because the faster cars will cross the slow vehicles in a relatively short period of time that will further destabilize the clusters. Similar to [24], for a vehicle to RSU communication, a reinforcement learning algorithm is used [25] to create clusters and fuzzy logic is used to make the clusters more stable considering vehicle mobility, vehicle distribution, and channel condition; however, the QoS is an issue when vehicle density grows faster.

*3.3.2  K-means with Fuzzy Logic*

Authors in [23-25] are using Q-learning algorithm where [26] addresses the issue of traffic congestion in a dynamic vehicle environment using k-means clustering algorithm integrated with

TABLE IV
EVALUATION OF HYBRID STRATEGIES

| Ref. | Algorithms | Evaluation parameters |
|---|---|---|
| [23] | Fuzzy, Q-learning | CH duration, percentage of stability, service delay |
| [24] | fuzzy, Q-learning | PDR, throughput, no. of handoffs, delay |
| [25] | Fuzzy, Q-learning | PDR, no. of collided frame, delay, throughput |
| [26] | K-means, fuzzy | Congestion |
| [27] | Fuzzy, Dolphin Swarm | Detection rate, detection time, false positive rate |

Arduino controller and a PHP web server. A fuzzy rule-based inference system is proposed considering four attributes: vehicle speed, rain, fog, and brake frequency. For all the vehicles, the fuzzy congestion output is sent to a PHP cloud server through an ESP8266 wi-fi module. This module also generates a two-dimensional position of the vehicles as an alternate of GPS. The PHP server uses K-means clustering algorithm to form the clusters without any assistance from RSUs. However, the time to connect to the PHP server and to send or receive information from an external server is required that can cause additional delay. Moreover, k-means always choose the centroid as the CH and for any change in the cluster may cause to change the CH every time.

*3.3.3  Dolphin Swarm Algorithms with Fuzzy Logic*

Authors in [11-25] use a single CH for a cluster where the CH acts as the leader of the cluster. To reduce the overload of the CH in the cluster, a multiple CH scheme is proposed in the Hybrid fuzzy multi-criteria decision making (HF-MCDM) [27] where fuzzy analytic hierarchy process and TOPSIS methods are combined together to form a cluster making the fuzzy decision optimal. The load of the leader of a cluster is distributed among the CHs. To secure the communication, intuition detection system has been proposed utilizing the Dolphin Swarm behavior instead of rule-based system to detect newer attacks which are not present in the database and to differentiate between the malicious and the normal nodes. The CH is selected based on velocity, social contact, integrity, availability, etc. Each CH will appoint another CH based on security and trustiness, the new CH will appoint another new CH, hence, a clustered swarm of dolphins are created. However, clustering efficiency or clustering stability issues are not described, and no simulation result provided to measure the clustering efficiency or stability of the clusters based on multiple CHs.

Table IV presents the VANET clustering algorithms which are based on fuzzy logic algorithms. The second column describes the machine learning algorithm that is used with fuzzy logic to create the hybrid strategies.

TABLE V
COMPARISON THE PERFORMANCES OF INTELLIGENT STRATEGIES

|  | Cluster formation | CH selection | Clustering efficiency | Clustering stability |
|---|---|---|---|---|
| Machine learning | √ |  | √ |  |
| Fuzzy logic |  | √ |  | √ |
| Hybrid | √ | √ | √ | √ |

*3.4 Summary of Intelligence-based Strategies*

The most important parts in VANET clustering process are the cluster formation and the CH selection. Efficiency of clusters largely depends on the cluster formation process where the stability of the clusters depends on the CH selection process. The efficiency of the clusters is evaluated in terms of packet loss, end-to-end delay and throughput more frequently while the stability of the clusters is evaluated based on average CH duration, average CM duration, number of clusters, and the number of CH changes. Table V shows a comparative analysis of intelligent VANET clustering strategies.

The stability of the clusters highly depends on the lifetime of the clusters. Generally, k-means-based algorithms [11-14] cannot provide high lifetime to the clusters, because if a single vehicle joins or leave the cluster, the entire algorithms is reset, and the CH can be changed frequently. Consequently, the stability is also affected. Hierarchical clustering algorithms [15] do not suffer this problem and can provide higher lifetime and higher stability compare to k-means algorithms. Fuzzy logic algorithms [18-27] consider the prediction of future movement that provide more flexibility compare to k-means. As a result, fuzzy logic algorithms can provide higher lifetime and more stability. Similarly, hybrid strategies can also provide higher lifetime and better stability compare to k-means algorithms.

Machine learning-based algorithms perform better in cluster formation process and can create efficient cluster. Machine learning-based algorithms [11, 12, 15, 23-25] are generally evaluated in terms of efficiency parameters such as PDR, delay, and throughput. However, due to vehicle mobility, clusters break frequently. Hence, along with clustering efficiency, clustering stability is also important for VANET clustering strategies. Fuzzy logic-based solutions can provide better stability by predicting future movement of the vehicles. These algorithms are generally evaluated in terms of CH and CM parameters such as CH duration, CM duration etc. [18-20, 22, 23]. In hybrid architecture [23-27], machine learning algorithms are used for cluster formation process to create efficient clusters while fuzzy logic is used to make the clusters more stable by selecting the most qualified vehicle as the CH, hence, learning process helps the clusters to learn from environment about the changes so that it can predict the movement of the vehicles and can make better decision to select the CH. Hence, cluster lifetime increases, as a result, cluster can have better stability.

## 4 MOBILITY BASED STRATEGIES

the literature can be classified into stability, routing, and QoS, etc. strategies according to the purpose of the algorithms.

In NEMO, MR can move from one access router to another access router along with its network retaining its IP address.

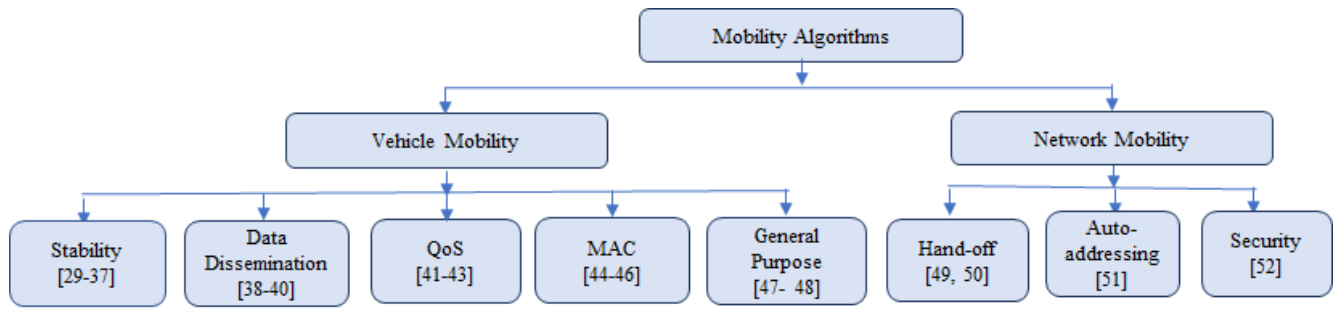

Fig. 9. Classification of mobility-based VANET clustering algorithms.

The most common clustering strategies in VANET is mobility-based strategies. The movement of the vehicle, such as relative speed, moving direction, acceleration, position etc., are the basic metrics used for mobility-based clustering. Due to the high mobility of the vehicles, clusters frequently break down in VANET. Therefore, instead of the efficiency of the clustering algorithms, stability of the clusters is the main concern in the vehicle mobility-based algorithms. The classification of mobility-based algorithms is presented in the Fig. 9.

In Section 4.1, vehicle mobility-based strategies are presented. Main purpose of the vehicle mobility-based clustering strategies is to provide more stability to the clusters; however, some mobility-based clustering strategies are proposed to facilitate data dissemination, MAC management, QoS, etc. Mobility algorithms are further divided based on stability, routing, QoS, MAC, and general-purpose clustering algorithms. One point is to note that the stability, routing, QoS, or MAC-based algorithms are not necessarily based on mobility algorithms only, rather mobility based algorithms presented in

Many efficient algorithms, such as [28], have been presented in the literature for efficient routing for NEMO. VANET clustering concept has high similarity with the concept of NEMO and some VANET clustering techniques are based on NEMO. Therefore, we classified NEMO clustering algorithms in a separate section (Section 4.2). This is a *difference* of our works with the existing works that we provided a distinct classification for the NEMO-based algorithms.

### 4.1 Vehicle Mobility Algorithms

The most popular clustering techniques developed for VANET are based on vehicle mobility. Even though the main purpose of the mobility-based clustering strategies is to provide stability to the clusters, some mobility-based clustering strategies are proposed to provide data dissemination, MAC management, Quality of Service (QoS), etc. Mobility metrics such as the average relative velocity of the vehicle, acceleration, position, direction, etc. are considered to select the CH and form the clusters. Stability-based parameters such as average CH duration, average CM duration, number of state change, etc. are evaluated for mobility-based clustering strategies.

### 4.1.1 Stability Algorithms

Dynamic Clustering Algorithm (DCA) [29] is proposed to increase the stability of the clusters in a highly dynamic environment. The cluster is formed considering the similarity of the vehicles in terms of relative speed. The CH is selected based on average velocity and acceleration of the vehicles without considering direction or future movement of the vehicle. Performance is evaluated based on two parameters only; average CH duration and average number of cluster change; however, average CM duration, average state change, etc. are not evaluated to measure the stability of the clusters. Similar cluster formation process and CH selection criteria are followed in [30-33]. Additionally, [33] prevents frequent merging of the clusters to increase the stability of the clusters. To accomplish this, several CHs are allowed to be present within the communication range for a certain amount of time. Hence, the lifetime of the clusters increases and increases stability. Like [29], CH lifetime is evaluated in [33], moreover, percentage of CH in relation to the total number of vehicles is also evaluated along with the number of state change, however, CM related parameters are neglected.

Goonewardene et al. [34] proposed a robust mobility-adaptive clustering (RMAC) where cluster formation and the CH is selected based on relative speed, location, and direction of the vehicle. Unlike other clustering strategies, each vehicle maintains a routing table for neighbor vehicles which are beyond its communication range. A vehicle can operate in a dual state, i.e., a vehicle can act as a CH and as a CM simultaneously. In a dual state, the vehicle will be the CH for its own cluster and a CM for one or more other clusters. CMs are one-hop clusters where all the CMs are within the communication range of the CH; however, not all the vehicle within the range of a CH are CM. Therefore, overlapping of the clusters is possible, and multiple CHs can operate in proximity without merging; however, stability is measured based on two parameters only: CM duration and re-clustering time, which is not sufficient to measure the performance of stability in a dynamic manner.

A mobility prediction-based clustering (MPBC) is proposed by Ni et al. [35] using the Doppler effect during the movement of the cars. To predict the relative speed, vehicles exchange Hello packets periodically and calculate Doppler shifts to initiate clustering process. The vehicle with the lowest relative speed is selected as the CH. Once the cluster is formed, vehicles exchange message to predict the future movement; however, an analytical model is presented comparing with two MANET clustering algorithms, no simulation result is presented to compare the strategy with a VANET clustering algorithm. Similarly, software-defined networking enabled social-aware clustering (SESAC) algorithm is proposed [36] to improve the cluster stability based on a social pattern. The moving pattern and sojourn time are considered to get the social pattern. Based on the historical movement pattern vehicles are grouped in a cluster who follow the same route. Relative speed and inter-vehicle distance are considered to select the CH. Even though simulation results presented in [36] shows that it can improve the performance from [35]; however, the strategy is evaluated based on cluster lifetime along with clustering overhead only. The lifetime of a cluster is an important parameter but cannot be the only measurement to measure the stability of the clusters. A cluster may have a longer lifetime, but frequent CM disconnection can decrease its stability, hence, CM related parameters should also be considered to measure the stability of the clusters.

Along with the location and direction, the speed difference is also considered in [30] to form a stable cluster, specifically for highway environment. The vehicles that show similar mobility patterns are clustered in a single cluster where the vehicles with high mobility are in a single cluster and the vehicle with low mobility form a different cluster. A suitability value is used to elect the CH using velocity, direction, location, and degree of node. Merging of clusters is allowed in this scheme when two CHs come closer with a relative speed less than a pre-defined threshold value; however, CM related parameters such as average CM duration and average state change are not evaluated. Along with the position and speed of the vehicles [30], acceleration is also considered in [31] to provide more stability and security, however, the scheme is optimized for highway only. Additionally, average CH duration achieved in this scheme is not significant which decreases cluster lifetime.

Traffic pattern of buses is used to improve stability in [32] by decreasing the number of CH change. Velocity, position, and direction of the vehicles are used as the mobility metric along with fixed-route pattern of buses in urban area. The number of CH changes is evaluated only; however, stability and lifetime of the clusters do not depend on a single parameter of the CH. Moreover, CM related parameters are ignored.

Moving direction, relative vehicle position, and link lifetime are considered to form a cluster in [37]. A temporary state for the CHs and a safe distance threshold have been introduced to increase the stability of the clusters. CH is chosen from the vehicles which is nearest to the center of a cluster so that its neighbor can spend more travel time to leave the cluster. Temporary cluster head (CHt) is used to begin the cluster formation process and it becomes CM if it has no member, otherwise, it changes to CH. If two clusters come closer than a predefined safe distance threshold, then they merge to become a single cluster. Along with the three parameters used in [30], four more parameters have been used to evaluate clustering stability: average CM duration, average state change rate per node, number of vehicles in clustered state, and CM disconnection frequency. However, this scheme is optimized for urban scenarios without considering the reliability issues, therefore, in a sparse environment this scheme creates a greater number of clusters that will decrease the average CH duration. Consequently, cluster stability will decrease, and cluster lifetime will become low. Moreover, in some cases, many vehicles can have no cluster to join.

*4.1.2 Data Dissemination/ Routing Algorithms*

In [38], a clustering-based data dissemination protocol is proposed improving a non-cluster-based routing protocol. The most reliable vehicle is selected as the CH based on the average relative velocity of the vehicles. The relative velocity, position, and direction are considered during cluster formation to reduce the disconnected problem during low density in highway and the broadcast storm problem during high density in urban area; clustering stability parameters such as CH duration, CM duration, etc. are not evaluated. Therefore, the lifetime of the clusters in this algorithm questionable.

A clustering-based routing algorithm is proposed in [39] to reduce control overhead. Location, direction, velocity, destination, etc. are considered to form the clusters and to select the CH; however, it suffers a frequent number of cluster changes that reduce the lifetime and stability of the clusters. Prediction-based routing protocol has been proposed in [40] for the medical vehicle in time of emergency to increase the reliability and stability of the clusters. Metrics such as medical vehicle attributes, road conditions, and driving environments are considered to predict a route for emergency vehicles to avoid high traffic. The present position, the predicted future position, id of the community, and relative distance are considered to form a cluster. The relative distance and the highest id are considered to select the CH; however, no mobility parameter or stability parameters are evaluated, hence, the lifetime of the clusters and the clustering stability are unmeasured.

*4.1.3   QoS Algorithms*

To provide QoS in the case of RSU failure, a concept of intelligent CH is introduced in [41] where density of the vehicles is considered along with distance and speed to select the intelligent CH. This concept can be used in any RSU-based clustering algorithms during RSU failure; however, stability parameters are ignored that can reduce the lifetime of the clusters. [42] also proposed a density-based scheme to reduce congestion and increase QoS using a trained dataset. Like [37], four states of the vehicles are considered where a supplementary CH state is used during cluster formation process which can be compared with the temporary CH state of [37]. If the node density crosses a predefined threshold, the cluster is formed. The most stable and reliable node is chosen as the CH; however, the performance is compared with a very old strategy, rather comparison with some of the new clustering techniques is required to establish the competency of the algorithm.

Mobility metrics with the QoS metrics such as bandwidth, the degree of the neighborhood, and link quality are considered in [43] and the CH is selected based on the suitability of these values. Clusters are divided into two layers: static clustering for V2I communication and dynamic clustering for V2V communication. When the CHs are in the communication range of the RSUs, all vehicles act as the CMs. When no RSU is reachable, a CH acts like a router. Merging of clusters is allowed if they reside within TR for a certain amount of time. Four parameters are evaluated: CH duration, number of clusters, packet delivery ratio (PDR), and clustering overhead; however, PDR or overhead can be better parameters to measure clustering efficiency rather than the stability of the clusters. Moreover, CM related parameters are not evaluated. Besides, the simulation results presented are only for highway scenarios.

*4.1.4   MAC Algorithms*

Clustering-based MAC protocol is proposed in [44-45] for faster delivery of safety messages, and in [46] for efficient resource management. [44] introduces three new control packets instead of RTS/CTS (request to send/clear to send) packet for the cluster formation process. When an isolated vehicle does not get any response, it becomes CH. If the number of vehicles is less than the delay will be the less and CH will broadcast the CM list to all CMs. [45] and [46] deal with the hidden terminal problem. [46] tries to reduce the packet conflict due to hidden terminal problem, and [45] tries to solve the hidden terminal problem using a reserve channel that can be used by safety messages even during the congestion. [45] uses velocity, and acceleration to form the cluster where [46] uses relative speed to form the cluster and to choose the CH. None of these algorithms [44-46] evaluated the stability parameter that can affect the lifetime of the clusters. These clustering-based MAC protocols are working as evidence that clustering not only solves the scalability problem but also can be utilized for various purposes in VANET.

### 4.1.5 General-purpose Algorithms

A general-purpose clustering algorithm is proposed in [47] to provide a more stable and efficient cluster considering velocity, position, direction, and link quality. Double-head cluster is used, so that a CM does not get disconnected from its cluster even it loses connection with the primary CH. Four states of vehicles are used like [37] where one vehicle acts as a mirror of the CH and works as the backup CH when a CM loses connection with its primary CH. Four states of vehicles are used like [37] where one vehicle acts as a mirror of the CH and works as the backup CH when a CM loses connection with its primary CH. The relative position in the cluster, relative speed, average signal-to-noise ratio, and average link expiration time are considered to become the CH; however, performance comparison with the recent clustering algorithms need to be presented. Also, how the two CHs handle PDR, delay, and throughput is not evaluated. A weighted clustering algorithm is proposed in [48] where the reputation of the vehicles is considered in the cluster formation process. The CH is selected based on the reputation of the vehicles along with direction, position, velocity, number of nearby vehicles, and lane ID. The reputation of a vehicle is calculated as a number, the vehicle worked as a CH. PDR, number of clusters, and control overhead are evaluated, however, the results are compared with two MANET algorithms, comparison with VANET clustering algorithms need to be evaluated.

Table VI summarizes the mobility based clustering algorithms. The two columns represent the purpose of the algorithms and the evaluation parameters respectively.

TABLE VI
EVALUATION OF VEHICLE MOBILITY-BASED STRATEGIES

| Ref. | Purpose/application | Evaluation parameters |
|---|---|---|
| [29] | Stability | CH lifetime, no. of cluster change |
| [30] | Stability | Cluster lifetime, no. of cluster change, no. of cluster |
| [31] | Stability, security | CH duration, CM duration, no. of state change, packet loss ratio |
| [32] | Stability | CH change |
| [33] | Stability | Cluster lifetime, percentage of CH, state change |
| [34] | Stability | CM duration, re-clustering time |
| [35] | Stability | CH duration, connection duration, re-association rate/time |
| [36] | Stability | Cluster lifetime, overhead |
| [37] | Stability | No. of clusters, CH/CM duration, CH/CM change rate, state change, efficiency |
| [38] | Data dissemination | Throughput, energy consumption, reliability, delay |
| [39] | Routing | CH duration, number of Ch change, PDR, delay, overhead |
| [40] | Selective routing | PDR, delay |
| [41] | QoS | Throughput, delay, PDR, packet loss |
| [42] | Congestion, QoS | CH/CM duration, number of clusters, PDR, delay |
| [43] | QoS | CH duration, no. of clusters, PDR, overhead |
| [44] | MAC, safety messages | Throughput, PDR, delay |
| [45] | MAC for safety message | Throughput, latency, overhead, packet loss |
| [46] | MAC | Throughput, delay, PDR |
| [47] | General-purpose | CH duration, CM duration, no. of state change. Overhead, single CH, single vehicle, no. of clusters |
| [48] | Weighted | PDR, number of clusters, overhead |

## 4.2 NEMO Algorithms

Some of the clustering algorithms adopted the NEMO concept in the VANET environment. NEMO-based clustering techniques mainly developed for faster handoff, i.e., to reduce the total number of handoffs and handoff latency. Table VII evaluates NEMO-based algorithms.

### 4.2.1 Handoff Algorithms

To reduce the number of handoffs as well as handoff latency, clustering strategy is applied in [49] for NEMO-based VANET. The MR is considered as the CH of the cluster. The MR and mobile nodes connected with the MR are treated as a cluster. Since CH handles the routing procedure in clustering strategies, vehicles are divided into clusters to minimize the number of handoffs. The vehicles acquire their CoA from the CMs of the new cluster prior to actual handoff, hence, latency can be reduced. However, no simulation result is provided comparing the result of the scheme to prove the effectiveness of the scheme in VANET scenario.

To solve the handoff and packet loss problem in high-speed VANET, NEMO-based protocol [50] is proposed for highway. The MR is the CH and the network is treated as the cluster. In this protocol, the car can acquire IP address from the VANET through a V2V communications to achieve network mobility. To execute the pre-handoff procedure, the vehicle relies on the assistance of the front vehicle to acquire its care-of address, or it may acquire its new IP address through multi-hop relays from the vehicle on the lanes of the same or opposite direction. Hence, it reduces the handoff delay and maintains the connectivity to the Internet; however, comparison with other clustering algorithms are absent.

### 4.2.2 Auto-addressing Algorithms

Mobile IPv6 (MIP) based dynamic auto-addressing protocols have been investigated in the VANET scenario in cluster-based addressing scheme (CBAS) [51]. The MR is the CH and the network is worked as the cluster. In this MIP-based scheme, incoming vehicles are assigned unique IP address and clustering is used to overcome the problem of maintaining the unique IP address since vehicle communicate through its CH.

TABLE VII
EVALUATION OF NEMO-BASED STRATEGIES

| Ref. | Algorithms | Evaluation |
|------|------------|------------|
| [49] | NEMO | Packet loss, delay |
| [50] | NEMO | Handoff latency, PDR, overhead |
| [51] | NEMO | IP address management |
| [52] | NEMO | Power, delay |

Vehicles are clustered based on their relative speed and the CH assigns the IP address to its member and ensures that the assigned IP addresses in its vicinity are unique. However, no simulation result is provided that can show the effectiveness of the scheme.

### 4.2.3 Security Algorithms

NEMO-based solution for VANET clustering discussed in [49-51] are mainly to solve the handoff problem, while [52] applies clustering for NEMO to increase the security for vehicular communication. The network is treated as the cluster while the MR is called the CH. In this scheme, vehicles are grouped in different clusters to reduce the probability of attack and different clusters can be accessible through their corresponding CH only that works as extra layer protection. However, like other NEMO-based schemes, performances are compared with NEMO-based solutions, no results are provided comparing the scheme with other VANET clustering algorithms.

*4.3 Summary of Mobility-based Strategies*

Most of the vehicle mobility-based solutions are proposed to increase the stability of the clusters, because even an efficient clustering algorithm can perform worse in the high mobile environment. For this reason, maintaining the stability of the clusters is given priority in mobility-based strategies. Some of the mobility-based strategies also serve data dissemination, MAC, QoS, etc. Mobility-based strategies have been evaluated mostly using average CH duration and average CM duration. Network mobility-based solutions used in VANET mainly for hand-off purpose; however, none of them compare their performances with other clustering algorithms such as machine-learning or fuzzy logic-based algorithms. Hence, the suitability of NEMO-based clustering protocol for the VANET environment is still uninvestigated.

The stability of the clusters depends on the lifetime of the clusters. Generally, mobility-based algorithms [29-48] consider mobility related parameters to form a cluster and to select the CH. In many cases, the mobility-based algorithms intentionally leave a better cluster to increase the stability of the clusters by increasing the lifetime of the clusters. NEMO-based algorithms [49-52] mainly concentrate on hand-off procedure and cannot provide higher lifetime and stability compare to vehicle mobility-based algorithms.

## 5 MULTI-HOP BASED STRATEGIES

Reducing the number of clusters is one of the challenges for VANET. Many clustering algorithms are published in the literature based on multi-hop transmission of the packet to reduce the number of clusters. Here, a CH can cover a larger area and can provide better stability. Our work is clearly *different* from the existing surveys [4-10] that we evaluated multi-hop algorithms in detail. In Section 5.1, 2-hop-based algorithms are presented, this section is limited to algorithms that are specifically mentioned to implement up to 2-hop or did not provide any result for more than 2-hop. Section 5.2 presents the algorithms which are not limited to 2-hop, i.e., algorithms can implement any number of hops. Fig. 10 shows the classification of multi-hop VANET clustering algorithms.

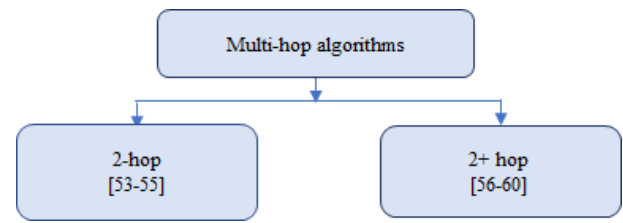

Fig. 10. Classification of multi-hop-based clustering.

*5.1 2-hop Algorithms*

In 2-hop communication, the CH can reach up to 2-hops of vehicle for its coverage. A clustering-based routing algorithm is proposed in [53] considering vehicle position and moving direction to form the cluster. Each vehicle broadcasts a beacon message including its latitude, longitude, and direction. The receiving vehicle will first check the number of hop count value of the received message, if the number of hops is larger than a threshold value, it will discard the message. Upon receiving the acknowledgment, the sender vehicle updates its routing table by calculating the Euclidean distance of the vehicles and the vehicles belong to its closest CH. Simulation results are provided evaluating PDR, routing overhead, etc.; however, end-to-end delay, throughput, etc. are not evaluated which were necessary to measure clustering efficiency. Because, in multi-hop communication, packet needs to travel longer distance compare to single-hop clustering algorithm, hence, in multi-hop communication end-to-end delay increases while throughput decreases.

In the absence of GPS, a multi-hop hierarchical clustering algorithm (HCA) is proposed in [54] to connect the vehicle into a two-hop cluster in the shortest possible time considering time and space complexity; compromising the quality of the cluster. The center vehicle is chosen as the CH. In these multi-hop schemes, if a CH loses its members and within the range of a CG of another CH, the single CH merges with the CH of the larger cluster. Additionally, if a CH arrives in the TR of another CH and the first CH has a shorter distance to the CMs compare to the second CH, both clusters will merge with the first CH as the new CH and the second CH as the CG. If necessary, the cluster can be optimized in the maintenance phase after creating the initial cluster; however, the simulation results are provided comparing the data with a MANET clustering technique in terms of number of clusters and number of cluster change [86], where time spent in cluster is also evaluated in [54]. No significant clustering stability or clustering efficiency-related parameters are evaluated.

Network criticality is used as the metric in a robust multi-hop-based algorithm, presented in [55]. In this criticality-based clustering (CCA) technique, the robustness of an undirected network graph to the change of the environment, such as the destination change or topology change, is termed as network criticality and interpreted as an electrical circuit where vehicles show resistant to any change in the environment. Two clusters can merge if they show a similar pattern and come close, to form a single cluster. Some parameters like number of clusters, average lifetime, cluster size, etc. have been evaluated; however, the important parameters like duration of vehicles spent as CH, duration of vehicles spent as CM and the number of state change per vehicle are not evaluated.

Table VIII represents 2-hop-based multi-hop algorithms with the evaluation parameters.

### 5.2  2+ hop Algorithms

In many multi-hop-based algorithms, CH can reach 2 or more hops, such as 3-hop, 4-hop, or 5-hop coverage.

Vehicles are allowed to broadcast beacon message periodically and calculate the relative mobility based upon two consecutive beacon messages received from the same node in [56]. Each vehicle calculates the aggregate mobility value, which is the sum of relative mobility values and the weight value for all the neighboring nodes in $N$-hops. The vehicle then broadcast their aggregate mobility value in the $N$-hop neighborhood and the vehicle with the smallest aggregate mobility value is selected as the CH. If a vehicle receives multiple beacon messages, it selects the CH which is the closest in terms of hop count. The vehicle with the lowest relative mobility is selected as the CH when more than one CH candidates have the same hop count. Average CH duration, average CM duration, and the number of CH change have been evaluated, however, the number of state change or number of vehicles in the clustered state are not evaluated which are also important parameters for clustering stability.

A distributed multi-hop clustering algorithm for VANETs based on neighborhood follow (DMCNF) is proposed in [57] where relative mobility is given preference. It considers the relationship among the vehicles within the neighborhood to choose the CH. Due to high mobility vehicles cannot identify the vehicles in its multi-hop neighbors, therefore, they consider the vehicle in one hop as a single cluster. CM chooses its CH based on the stability of the vehicles and their history of the movement which is denoted as neighborhood follow relationship. All the CMs follow the CH. They do not use location service rather depend on the topology. Performance is measured in terms of CH duration, CM duration, number of CH change, number of clusters, and overhead.

TABLE VIII
EVALUATION OF 2-HOP ALGORITHMS

| Ref. | No. of hop | Evaluation parameters |
|------|-----------|----------------------|
| [53] | 2-hop | PDR, routing overhead |
| [54] | 2-hop | No. of CH, cluster change, time spent in cluster |
| [55] | 2-hop | No. of clusters, average lifetime, cluster size, CH/CM change time |

When the relative speed of the CH changes, it causes frequent CH change. To increase the routing performance by reducing the number of CH change, a passive multi-hop clustering (PMC) algorithm is proposed in [58] by improving [53]. The number of candidates to become the CH in a multi-hop scenario is more than a single-hop clustering and the most stable vehicle becomes the CH. In a multi-hop clustering, the CH can have N-hop coverage compare to single-hop clustering and can achieve more stability and reliability. Cluster merging is allowed in this scheme to reduce the number of clusters. At the same time, merging is allowed only two CHs overlap for a certain amount of time so that stability also increases. Simulation results are provided considering CH duration, CM duration, number of CH change, and overhead. An RSU assisted multi-hop scheme is proposed in [59] based on [53]. A new vehicle broadcasts hello packet to all its neighbors with its position, speed, and direction, and the neighbors reply with another hello packets that increase the number of packet dissemination and the number of packet loss. To solve this problem, the new node communicates with the RSU to receive information about the stability of the clusters and can join into the cluster in a relatively faster time, however, important parameters are not evaluated.

Vehicular multi-hop algorithm for stable clustering (VMaSC) was proposed in [60]. Cellular technologies have been used in conjunction with IEEE 802.11p to reduce the cost of communications between vehicles and base stations as well as the number of handoffs. Average relative speed is measured among the neighbors of the *N*-hop to create the clusters and the vehicle with the lowest mobility wins to become the CH. A new vehicle adds to the neighbor CH or CM in a multi-hop manner instead of connecting with the CH directly. Merging is allowed in this scheme when two CHs overlap for a certain amount of time. In a multi-hop communication, the CH acts as a dual-interface node where CH communicates with CMs via IEEE 802.11p interface and connects the cluster to the cellular network via the LTE interface; however, simulation results are not impressive for V2V communication. Even though RSU assisted communication can perform better, in the absence of RSU, a part of the proposed scheme would not work.

TABLE IX
EVALUATION OF 2+ HOP ALGORITHMS

| Ref. | No. of hops | Evaluation |
|---|---|---|
| [56] | 2-hop, 3-hop, 5-hop | CH/CM duration, no. of CH change |
| [57] | 3-hop | CH/CM duration, no. of CH change, no. of clusters, overhead |
| [58] | 3-hop | Overhead, no. of packet/time for cluster selection |
| [59] | 3-hop | CH/CM duration, CH change, overhead |
| [60] | 2-hop, 3-hop | CH/CM duration, CH change rate, no. of un-clustered vehicle, overhead |

Table IX represents 2+ hop multi-hop algorithms where the numbers of hops are mentioned with the evaluation parameters of the algorithms.

### 5.3 Analysis on Multi-hop-based Strategies

Multi-hop-based algorithms cover a larger area compare to single-hop clustering algorithm; hence, multi-hop algorithms create a lower number of clusters for the equal number of vehicles. On the other hand, clustering overhead increases in the multi-hop algorithms because of the increased number of messages. Moreover, performance of the clustering efficiency parameters such as PDR, latency, throughput is not as good as the clustering stability parameters for multi-hop strategies.

The stability and reliability of the clusters generally depends on the lifetime of the clusters. Average lifetime of a cluster in the multi-hop strategies [53-60] are higher than the average lifetime of a cluster in a single-hop strategy. Because, instead of creating a new cluster, vehicles join into an existing cluster; however, member vehicle addition and deletion also happens more frequently. For this reason, machine learning-based algorithms, such as k-means, are not used in multi-hop-based strategies, since any member of addition or deletion changes the entire dynamic of the clusters. Unlike k-means, multi-hop-based strategies are generally reluctant to the change of the CH of the clusters.

## 6 CHALLENGES AND FUTURE DIRECTIONS

The most important parts in VANET clustering are the cluster formation and the CH selection process. CMs join the cluster in the cluster formation process. Clustering algorithms in VANET are dynamic in nature and logically applied in application level. No vehicle changes its position physically based on clustering, rather the vehicles join into the cluster based on its physical position. Joining in a cluster is an optional choice for the vehicles. Efficiency of clusters largely depends on the cluster formation process where stability of the clusters depends on the CH selection process. The efficiency of the clusters is evaluated in terms of packet loss, end-to-end delay, and throughput more frequently while the stability of the clusters is evaluated based on average CH duration, average CM duration, number of clusters, and number of state changes, etc.

Machine learning and fuzzy logic algorithms have been evaluated in terms of packet loss [11, 12, 15, 19, 21, 24, 25], end-to-end delay [12, 15, 19, 21, 24, 25], throughput [11, 12, 15, 21, 24, 25] more frequently while mobility-based and multi-hop-based algorithms are evaluated in terms of average CH duration, average CM duration, number of clusters, number of changes, etc. If different combinations are possible for a given scenario where the most efficient path can be established but will remain for a few seconds where the second-most efficient path can be established with a better lifetime. Intelligent-based strategies generally create the first scenario where mobility-based and multi-hop-based strategies choose the second method. Therefore, it can be concluded that intelligent algorithms are concentrating on efficiency while mobility and multi-hop strategies emphasize on stability of the clusters.

*Intelligence-based* clustering algorithms utilize machine learning algorithm and fuzzy logic to create clusters and to select the CH. Machine learning-based algorithms perform better in cluster formation process and can create efficient cluster. However, due to vehicle mobility, clusters break or change frequently. Hence, along with clustering efficiency, clustering stability is also important for VANET clustering strategies. Fuzzy logic-based solutions can provide better stability compare to machine learning algorithms predicting the future movement of the vehicles. Therefore, hybrid architecture of machine learning and fuzzy logic algorithms combine machine learning algorithms with fuzzy logic to create efficient and stable clusters. In hybrid architecture, machine learning algorithms is used for cluster formation process to create efficient clusters while fuzzy logic is used to make the clusters more stable by selecting the most qualified vehicle as the CH. Hence, learning process of the vehicles helps the clusters to learn from environment about the changes and triggers appropriate action based on the changes.

Moreover, we did not find any intelligence-based multi-hop clustering. Indeed, multi-hop clustering can cover larger area compare to single-hop clustering and can reduce number of clusters. Therefore, the existing approaches can be extended from single hop to multi-hop clustering to reduce the number of clusters.

*Mobility-based* clustering approaches are the most common technique for clustering in VANET where relative speed and mobility pattern get importance. Although vehicle mobility is the key point for mobility-based schemes, some research works performed clustering based on network mobility considering the similarities between the clustering concept in VANET with NEMO. However, NEMO can be suitable for cellular architecture rather than an 802.11p environment because of its IP-based nature.

To provide more stability and reliability to the clusters, in *multi-hop-based* algorithms CH can get a larger coverage in a multi-hop manner and can reduce the number of clusters and number of CHs. Multi-hop algorithms can cover up to 2-hop, 3-hop, 4-hop, and 5-hop in the literature. One of the challenges for the multi-hop approaches is, when the number of hop increases, packet loss also increases. Moreover, multi-hop algorithms increase number of hops for packet transmission and increase end-to-end delay for the packets to reach from the source to the destination.

# 7 CONCLUSION

Detailed analysis of VANET clustering strategies is presented in this paper from intelligence, mobility, and multi-hop perspective with an intensive discussion on machine learning-based strategies, fuzzy logic-based strategies, hybrid strategies, mobility strategies, NEMO strategies, and multi-hop strategies. According to our findings, machine learning-based algorithms can create efficient cluster but cannot provide clustering stability because the clusters break frequently. Fuzzy logic-based algorithms can provide better stability compared to machine learning algorithms predicting the future movement of the vehicle. As a result, hybrid algorithms who combine machine learning algorithms with fuzzy logic algorithms can provide better stability compared to machine learning algorithms. On the other hand, mobility-based strategies consider stability-based parameters to form the cluster and to select the CH. These algorithms want to create more stable clusters sacrificing efficiency. NEMO-based approaches create the cluster using IP-based solution for VANET even though are not yet practically feasible. We have also found that no intelligent multi-hop clustering is presented in the literature. Therefore, the existing approaches can be extended from single hop to multi-hop algorithms to provide more stability to the clusters. Multi-hop strategies create extra overhead and sometimes create large size clusters that increases end-to-end delay for the packet delivery but can reduce the number of clusters. We can conclude that clustering in VANET is an open research issue since an efficient and stable clustering algorithm is still under research.